
\input phyzzx
%
\catcode`@=11 
\def\space@ver#1{\let\@sf=\empty \ifmmode #1\else \ifhmode
   \edef\@sf{\spacefactor=\the\spacefactor}\unskip${}#1$\relax\fi\fi}
\def\attach#1{\space@ver{\strut^{\mkern 2mu #1} }\@sf\ }
\newtoks\foottokens
%
%
%
%
%
%
%
%
%
%
\newcount\referencecount     \referencecount=0
\newif\ifreferenceopen       \newwrite\referencewrite
\newtoks\rw@toks
\def\NPrefmark#1{\attach{\scriptscriptstyle [ #1 ] }}
\let\PRrefmark=\attach
\def\refmark#1{\relax\ifPhysRev\PRrefmark{#1}\else\NPrefmark{#1}\fi}
\def\refend{\refmark{\number\referencecount}}
\newcount\lastrefsbegincount \lastrefsbegincount=0
\def\refsend{\refmark{\count255=\referencecount
   \advance\count255 by-\lastrefsbegincount
   \ifcase\count255 \number\referencecount
   \or \number\lastrefsbegincount,\number\referencecount
   \else \number\lastrefsbegincount-\number\referencecount \fi}}
\def\refch@ck{\chardef\rw@write=\referencewrite
   \ifreferenceopen \else \referenceopentrue
   \immediate\openout\referencewrite=referenc.texauxil \fi}
%
{\catcode`\^^M=\active 
  \gdef\obeyendofline{\catcode`\^^M\active \let^^M\ }}%
%
{\catcode`\^^M=\active 
  \gdef\ignoreendofline{\catcode`\^^M=5}}
{\obeyendofline\gdef\rw@start#1{\def\t@st{#1} \ifx\t@st\blankend%
\endgroup \@sf \relax \else \ifx\t@st\bl@nkend \endgroup \@sf \relax%
\else \rw@begin#1
\backtotext
\fi \fi } }
{\obeyendofline\gdef\rw@begin#1
{\def\n@xt{#1}\rw@toks={#1}\relax%
\rw@next}}
\def\blankend{}
{\obeylines\gdef\bl@nkend{
}}
\newif\iffirstrefline  \firstreflinetrue
\def\rwr@teswitch{\ifx\n@xt\blankend \let\n@xt=\rw@begin %
 \else\iffirstrefline \global\firstreflinefalse%
\immediate\write\rw@write{\noexpand\obeyendofline \the\rw@toks}%
\let\n@xt=\rw@begin%
      \else\ifx\n@xt\rw@@d \def\n@xt{\immediate\write\rw@write{%
        \noexpand\ignoreendofline}\endgroup \@sf}%
             \else \immediate\write\rw@write{\the\rw@toks}%
             \let\n@xt=\rw@begin\fi\fi \fi}
\def\rw@next{\rwr@teswitch\n@xt}
\def\rw@@d{\backtotext} \let\rw@end=\relax
\let\backtotext=\relax

\newdimen\refindent     \refindent=30pt
\def\refitem#1{\par \hangafter=0 \hangindent=\refindent \Textindent{#1}}
\def\REFNUM#1{\space@ver{}\refch@ck \firstreflinetrue%
 \global\advance\referencecount by 1 \xdef#1{\the\referencecount}}
\def\refnum#1{\space@ver{}\refch@ck \firstreflinetrue%
 \global\advance\referencecount by 1 \xdef#1{\the\referencecount}\refend}

\def\REF#1{\REFNUM#1%
 \immediate\write\referencewrite{%
 \noexpand\refitem{#1.}}%
\begingroup\obeyendofline\rw@start}
\def\ref{\refnum\?%
 \immediate\write\referencewrite{\noexpand\refitem{\?.}}%
\begingroup\obeyendofline\rw@start}
\def\Ref#1{\refnum#1%
 \immediate\write\referencewrite{\noexpand\refitem{#1.}}%
\begingroup\obeyendofline\rw@start}
\def\REFS#1{\REFNUM#1\global\lastrefsbegincount=\referencecount
\immediate\write\referencewrite{\noexpand\refitem{#1.}}%
\begingroup\obeyendofline\rw@start}
\newif\ifmref  
\newif\iffref  
\def\xrefsend{\xrefmark{\count255=\referencecount
\advance\count255 by-\lastrefsbegincount
\ifcase\count255 \number\referencecount
\or \number\lastrefsbegincount,\number\referencecount
\else \number\lastrefsbegincount-\number\referencecount \fi}}
\def\xrefsdub{\xrefmark{\count255=\referencecount
\advance\count255 by-\lastrefsbegincount
\ifcase\count255 \number\referencecount
\or \number\lastrefsbegincount,\number\referencecount
\else \number\lastrefsbegincount,\number\referencecount \fi}}
\def\xREFNUM#1{\space@ver{}\refch@ck\firstreflinetrue%
\global\advance\referencecount by 1
\xdef#1{\xrefend}}
\def\xrefend{\xrefmark{\number\referencecount}}
\def\xrefmark#1{[{#1}]}
\def\xRef#1{\xREFNUM#1\immediate\write\referencewrite%
{\noexpand\refitem{#1 }}\begingroup\obeyendofline\rw@start}%
\def\xREFS#1{\xREFNUM#1\global\lastrefsbegincount=\referencecount%
\immediate\write\referencewrite{\noexpand\refitem{#1 }}%
\begingroup\obeyendofline\rw@start}
\def\rrr#1#2{\relax\ifmref{\iffref\xREFS#1{#2}%
\else\xRef#1{#2}\fi}\else\xRef#1{#2}\xrefend\fi}
\def\multref#1#2{\mreftrue\freftrue{#1}%
\freffalse{#2}\mreffalse\xrefsend}
\def\doubref#1#2{\mreftrue\freftrue{#1}%
\freffalse{#2}\mreffalse\xrefsdub}
\referencecount=0
\def\refbreak{\hfil\penalty200\hfilneg}
\def\paperstyle{\papers}
\paperstyle   
%
%
%
\def\slacpub{\afterassignment\slacp@b\toks@}
\def\slacp@b{\edef\n@xt{\Pubnum={CERN--TH.\the\toks@}}\n@xt}
\let\pubnum=\slacpub
\expandafter\ifx\csname eightrm\endcsname\relax
      \fi

\font\seventeencp=cmcsc10 scaled\magstep3

\newif\ifCONF \CONFfalse
\newif\ifBREAK \BREAKfalse
\newif\ifsectionskip \sectionskiptrue

%
%
%
%
\def\NuclPhysProc{
\let\lr=L
\hstitle=8truein\hsbody=4.75truein\fullhsize=21.5truecm\hsize=\hsbody
\hstitle=8truein\hsbody=4.75truein\fullhsize=20.7truecm\hsize=\hsbody
\output={
  \almostshipout{\leftline{\vbox{\makeheadline
  \pagebody\makefootline}}}\advancepageno
     }
\def\papersize{\SIZE\OFFSET\skip\footins=\bigskipamount}
\def\SIZE{\hsize=10.0truecm\vsize=27.0truecm}
\def\OFFSET{\voffset=-1.4truecm\hoffset=-2.40truecm}
\def\makeheadline{
\iffrontpage\line{\the\headline}
             \else\vskip .0truecm\line{\the\headline}\vskip .5truecm \fi}
\def\makefootline{\iffrontpage\vskip  0.truecm\line{\the\footline}
               \vskip -.15truecm\line{\the\date\hfil}
              \else\line{\the\footline}\fi}
\paperheadline={\hfil}
\paperstyle}
%
%
%
%

\paperstyle
%
%
%
%
\def\ReprintVolume{\smallsize
\def\papersize{\hsize=18.0truecm\vsize=25.1truecm\voffset -.73truecm
    \hoffset -.65truecm\skip\footins=\bigskipamount
    \normaldisplayskip= 20pt plus 5pt minus 10pt}
\paperstyle\baselineskip=.425truecm\parskip=0truecm
\def\makeheadline{
\iffrontpage\line{\the\headline}
             \else\vskip .0truecm\line{\the\headline}\vskip .5truecm \fi}
\def\makefootline{\iffrontpage\vskip  0.truecm\line{\the\footline}
               \vskip -.15truecm\line{\the\date\hfil}
              \else\line{\the\footline}\fi}
\paperheadline={
\iffrontpage\hfil
               \else
               \tenrm\hss $-$\ \folio\ $-$\hss\fi    }
\def\sectionfont{\bf}    }
%
%
%
%
\def\SIZE{\hsize=15.73truecm\vsize=23.11truecm}
\def\OFFSET{\voffset=0.4truecm\hoffset=0.0truecm}
\def\papersize{\SIZE\OFFSET\skip\footins=\bigskipamount
\normaldisplayskip= 30pt plus 5pt minus 10pt}
\def\CERN{\address{{\sl CERN, 1211 Geneva 23, Switzerland\
\phantom{XX}\ }}}
\Pubnum={\rm CERN$-$TH.\the\pubnum }
\def\title#1{\vskip\frontpageskip\vskip .50truein
     \titlestyle{\seventeencp #1} \vskip\headskip\vskip\frontpageskip
     \vskip .2truein}
\def\author#1{\vskip .27truein\titlestyle{#1}\nobreak}

\def\p@bblock{\begingroup \tabskip=\hsize minus \hsize
   \baselineskip=1.5\ht\strutbox \topspace-2\baselineskip
   \halign to\hsize{\strut ##\hfil\tabskip=0pt\crcr
   \the \Pubnum\cr}\endgroup}
\def\makefootline{\iffrontpage\vskip .27truein\line{\the\footline}
                 \vskip -.1truein\line{\the\date\hfil}
              \else\line{\the\footline}\fi}
\paperfootline={\iffrontpage
 \the\Pubnum\hfil\else\hfil\fi}
\paperheadline={
\iffrontpage\hfil
               \else
               \twelverm\hss $-$\ \folio\ $-$\hss\fi}
%
%
\def\nup#1({\refbreak\ Nucl.\ Phys.\ $\underline {B#1}$\ (}
\def\plt#1({\refbreak\ Phys.\ Lett.\ $\underline  {#1}$\ (}
\def\cmp#1({\refbreak\ Commun.\ Math.\ Phys.\ $\underline  {#1}$\ (}
\def\prp#1({\refbreak\ Physics\ Reports\ $\underline  {#1}$\ (}
\def\prl#1({\refbreak\ Phys.\ Rev.\ Lett.\ $\underline  {#1}$\ (}
\def\prv#1({\refbreak\ Phys.\ Rev. $\underline  {D#1}$\ (}
\def\und#1({            $\underline  {#1}$\ (}
%
%

\def\rB{\hfil\penalty1000\hfilneg}
%
%
\hyphenation{sym-met-ric anti-sym-me-tric re-pa-ra-me-tri-za-tion
Lo-rentz-ian a-no-ma-ly di-men-sio-nal two-di-men-sio-nal}
%
%
%
%

%


%

%
\def\boxit#1{\vbox{\hrule\hbox{\vrule\kern3pt
\vbox{\kern3pt#1\kern3pt}\kern3pt\vrule}\hrule}}
\catcode`@=12
\paperstyle
\def\GoOl{\rrr\GoOl{
P.~Goddard and D.~Olive, in {\it Vertex Operators in Mathematics
and Physics}, Springer Verlag, (1984).}}
\def\GHMR{\rrr\GHMR{D.~Gross, J.~Harvey, E.~Martinec and R.~Rohm,
                    \nup256 (1985) 253;\rB  \nup267 (1986) 75;\rB
                                            \prl54 (1985) 502.}}
\def\GSWe{\rrr\GSWe{M.~Green, J.~Schwarz
and P.~West, \nup254 (1985) 327.}}
\def\ScW {\rrr\ScW {A.N.~Schellekens
and N.P.~Warner, \plt B177 (1986) 317;
\plt B181 (1986) 339;     \nup287 (1987) 317.}}
\def\GrS {\rrr\GrS {M.~Green and J.~Schwarz, \plt 149B (1984) 117.}}
\def\DiH{\rrr\DiH{
L.~Dixon and J.~Harvey, \nup274 (1986) 93.}}
\def\LLSA{\rrr\LLSA{
W.~Lerche, D.~L\"ust and A.N.~Schellekens, \plt B181 (1986) 71;
Erratum, \plt B184 (1987) 419.}}
\def\Sch{\rrr\Sch{
A.N.~Schellekens,
\plt B199 (1987) 427.}}
\def\KaPe{\rrr\KaPe{
V.G.~Kac and D.H.~Peterson, Adv. in Math. 53 (1984) 125.}}
\def\Nie{\rrr\Nie{
H.~Niemeier, Journal of Number Theory \und 5 (1973) 142.}}
\def\BMAP{\rrr\BMAP{
A.~Casher, F.~Englert, H.~Nicolai and A.~Taormina,
\plt 162B (1985) 121; \rB
F.~Englert, H.~Nicolai and A.N.~Schellekens, \nup274 (1986) 315; \rB
H.~Nicolai and A.N.~Schellekens, in {\it Proceedings of the 5th
Adriatic Meeting on Superstrings, Anomalies and Unification},
Dubrovnik (1986).}}
\def\AGMV{\rrr\AGMV{L.~Alvarez-Gaum\'e, P.~Ginsparg, G.~Moore and
                    C.~Vafa, \plt B171 (1986) 155.}}
\def\LNSW{\rrr\LNSW{W.~Lerche, B.E.W.~Nilsson, and A.N.~Schellekens,
\nup289 (1987) 609; \rB W.~Lerche, B.E.W.~Nilsson, A.N.~Schellekens and
N.P.~Warner, \nup299 (1988) 91.}}
%
\def\KLTE{\rrr\KLTE{H.~Kawai, D.~Lewellen and S.~Tye,
\prv34 (1986) 3794.}}
\def\ScYc{\rrr\ScYc{A.N.~Schellekens and S.~Yankielowicz,
\plt B226 (1989) 285.}}
\def\BB  {\rrr\BB  {L.~Dixon, P.~Ginsparg and J.~Harvey,
\cmp119 (1988) 221.}}
\def\God {\rrr\God {P.~Goddard, {\it Meromorphic Conformal Field Theory},
preprint DAMTP-89-01.}}
\def\ZamA{\rrr\ZamA{A.~Zamolodchikov, Theor.\ Math.\ Phys.\
\und 65 (1986) 1205.}}
\def\Gin {\rrr\Gin {P.~Ginsparg, {\it Informal String Lectures},
 to appear in the Proceedings of the U.K.~Institute for Theoretical
 High Energy Physics Cambridge 16 Aug.~- 5 Sept.~1987,   Harvard preprint
 HUTP-87/A077 (1987).}}
\def\BeBT{\rrr\BeBT {B. Gato-Rivera and A.N. Schellekens,
{\it Complete Classification of Simple Current Modular Invariants
for $({\bf Z}_p)^k$}, CERN-TH.6056/91,
to appear in Comm. Math. Phys.}}
\def\DGM {\rrr\DGM  {L. Dolan, P. Goddard and P. Montague,
\plt B236 (1990) 165.}}
\def\Ver {\rrr\Ver  {D. Verstegen, \nup 346 (1990) 349;
\cmp 137 (1991) 567.}}
\def\C{{\cal C}}
\hfuzz= 6pt
\def\Zbf{{\bf Z}}

\pubnum={6325/91}
\date{November 1991}
\pubtype{CRAP}
\titlepage
\title{Classification of Ten-dimensional Heterotic Strings}
\author{A. N. Schellekens}
\vskip 0.3truein
\CERN
\vskip 0.6truein
\abstract
Progress towards the classification of
the meromorphic $c=24$ conformal field theories is reported.
It is shown if such a theory has any
spin-1 currents, it is either the Leech lattice CFT, or it can be
written as a tensor product of Kac-Moody algebras with total central
charge 24. The total number of combinations of
Kac-Moody algebras for which meromorphic $c=24$ theories may exist
is 221. The next step towards classification is to obtain
all modular invariant combinations of Kac-Moody characters.
The presently available results are sufficient
to obtain a complete list
of all ten-dimensional heterotic strings.
Furthermore there are
strong indications for the existence of several
(probably at least 20) new meromorphic $c=24$ theories.
\endpage
\chapternumber=0
\pagenumber=1
The purpose of this paper is to answer a question that arose five
years ago. At that time nine heterotic string theories were known in
ten dimensions: the supersymmetric $E_8\times E_8$ and $O(32)$
theories \GHMR,
the tachyon-free non-supersymmetric $O(16)\times O(16)$ theory, and six
other non-supersymmetric theories with tachyons \multref\DiH{\AGMV\KLTE}.
Of these nine theories, eight have a rank 16 gauge group, and one
has a gauge group of rank eight. The gauge groups of the former theories
are embedded in a level 1, simply
laced Kac-Moody algebra, whereas the latter
theory has a Kac-Moody algebra $E_{8,2}$ (here and in the following
we denote the level by a second subscript).
All of these theories can be constructed using free fermions, and
furthermore they are the only ten-dimensional heterotic strings that
can be obtained this way. However, this does not prove that no other
ten-dimensional heterotic string theories can exist. In less than 10
dimensions one can certainly not write all heterotic string theories in
terms of free fermions. The purpose of this paper is to prove in a
construction-independent way that the list of ten-dimensional
heterotic string
theories that we know since 1986 is indeed complete.

The proof is valid for the class of ten-dimensional
critical heterotic strings satisfying the usual consistency
conditions. Such a theory is defined by a
conformal field theory consisting of
a right-moving NSR sector and a
left-moving bosonic sector. The right-moving
super-conformal field theory is
completely fixed by the requirement that the target-space has ten flat
space-time dimensions.
In the left-moving sector an ``internal''
$c=16$
conformal field theory $\C_{16}$
remains undetermined. Further restrictions
arise from one-loop modular invariance. This imposes the condition
the four spin-structures of the right-moving
world-sheet fermions must be combined
in a modular invariant way
with representations of the left-moving internal $c=16$ theory.
This must be done in such a way that all space-time states have the
correct spin-statistics relation, and that the ground state has
multiplicity 1.

An extremely useful tool for addressing superstring classification
problems is the bosonic string map \BMAP. This
map replaces the NSR sector (including superghosts)
of a heterotic (or type-II) string by a
bosonic sector, in such a way that a modular invariant heterotic string
satisfying the spin statistics condition is mapped to a modular
invariant\foot{The fact
that one-loop modular invariance is preserved was shown
in the last paper of \BMAP. A generalization of this
map to higher genus surfaces can be found in \Sch.}bosonic
string with positive multiplicities for all states.
In $d$ dimensions,
the light-cone $SO(d-2)$ algebra generated by the NSR fermions is
replaced by an $SO(d+6)\times E_8$ Kac-Moody algebra in the bosonic
theory.
Then a classification of all
ten-dimensional heterotic strings is equivalent to a classification
of all ten-dimensional bosonic strings with a right-moving Kac-Moody
algebra $D_{8,1}\times E_{8,1}$. The advantage of this map is
that spin-statistic signs, as well as signs in the modular
transformation
of the world-sheet spin-${3\over2}$
determinant are automatically taken care
of.

It has been shown in
\LLSA\ that the latter classification problem can be mapped
to yet another classification problem, by dropping the $E_8$ factor and
replacing the right-moving
$D_8$ factor by a left-moving one, whose conjugacy classes are paired
with the representations of $\C_{16}$ in exactly the same way. It is
easy to see that this map preserves modular invariance. The result is
a purely left-moving, unitary
CFT which is modular invariant by itself, built out
of representations of the tensor product $D_{8,1}\times \C_{16}$.
This theory
has central charge $c=24$, and, since it is modular invariant by
itself, only a single primary field with respect to the appropriate
chiral algebra (which is some extension of the chiral algebras of the
factors $D_{8,1}$ and $\C_{16}$). Conformal field theories with only
one primary field must have a central charge that is a multiple of 8,
and are sometimes called meromorphic CFT's \God.

This observation was used in \LLSA\ to show that the list of
ten-dimensional heterotic strings with a rank-16 gauge group is complete.
If $\C_{16}$ contains a Kac-Moody algebra of rank 16, it can be
written entirely in terms of free bosons, and the same is then true
for the tensor product $\C_{24}=D_{8,1}\times \C_{16}$. Hence $\C_{24}$
is given by a 24-dimensional even self-dual lattice, and since all
such lattices have been classified, the problem is solved.
All solutions can be obtained by considering all possible
embeddings of $D_{8,1}$ in the Kac-Moody algebras that
appear in the list of Niemeier lattices \Nie.
The $E_{8,2}$ theory cannot be obtained in
this way, but one can invert the argument to conclude that there
must exist a meromorphic $c=24$ theory built out of the
tensor product $B_{8,1} \times E_{8,2}$ \ScYc.
Conversely, if a list of
all meromorphic $c=24$ theories were available,
one could read off from it
all ten-dimensional heterotic strings by considering all
embeddings of $D_{8,1}$ in the Kac-Moody algebras of those theories.

Unfortunately, unlike the Niemeier lattices, a classification of the
more general meromorphic $c=24$ CFT's was not available so far. As a
classification problem this is of considerable
interest in its own right. Of
course one should keep in mind that
the $c=24$ theories are only the third member of an infinite
series $c=8k$, and that the number of free bosonic solutions
(\ie\ even self-dual lattices) increases extremely rapidly with $k$.
Nevertheless, the value $c=24$ is special, being the
smallest value for which the list of solutions is interesting, and
the largest for which a listing is practically possible. Furthermore
the $c=24$ theories have intriguing connections with other
mathematical concepts, such as the monster group. Finally, as the
second example in \ScYc\ shows, from a list of meromorphic $c=24$
theories we can get some information about solutions to another
interesting and unsolved problem, the classification of modular
invariants of Kac-Moody algebras.

The real purpose of this paper is to report progress towards the
classification of the $c=24$ meromorphic theories. Although this
classification is not yet finished, enough of it is now known to
carry out the enumeration of the ten-dimensional heterotic strings,
along the lines explained above.

The basic idea is to use the results of \ScW\ on the relation
between modular invariance and chiral anomaly cancellation in
effective field theories of string theories\rlap.\foot{In \ScW\ the
character valued partition functions used to prove
Green-Schwarz factorization \doubref\GrS\GSWe\
of anomalies in string theory were
explicitly constructed for theories constructed out of free bosons
or complex fermions. The generalization to higher level
Kac-Moody algebras
or non-simply laced ones, using the Weyl-Kac character formula, was
found later by Ginsparg, Moore and Vafa, and is described in
\Gin.}Note that this argument can be used to show
that the list of {\it supersymmetric}
heterotic string theories is complete.
This sort of argument is far less powerful for the
non-supersymmetric
string theories for two reasons: first of all there is no constraint
on the size of the gauge group because there are no gaugino
contributions to the gravitational anomaly, and
secondly one can have cancelling contributions to the chiral anomaly
due to Weyl fermions with opposite chirality.

However, the cancellation of chiral anomalies is only a small part of
the information contained in the ``character  valued partition function''
constructed in \ScW. Rather than trying to get constraints
on the $d=10$ heterotic strings or $c=24$ CFT's by applying
the condition of anomaly cancellation in some string theory, it turns
out to be much more effective to study directly the character valued
partition function of the $c=24$ CFT's.

A character-valued partition function
can be written down
if there is at least one spin-1 operator $J$ in
the theory. It can be shown \ZamA,
(and is anyhow not surprising) that
such operators either generate
Kac-Moody algebras (by definition non-abelian)
or $U(1)$'s, \ie\ their operator products are  restricted to the
following form
$$ J^a(z)J^b(w)={k\delta^{ab}\over (z-w)^2} +{1\over z-w}
i f^{abc} J^c(w) +\hbox{ finite terms} \ .\eqn\KacMoody $$
Then
one can organize the states at any level according  to
representations of the Lie-algebras generated by the
zero-modes
of the currents $J^a$.
The character-valued partition function has the
form
$$ \eqalign{P(q,F)&=\Tr e^{F \cdot J_0} q^{L_0 - {c/24}}  \  \cr
         &=\sum_{n=-1}^{\infty} q^n \Tr e^{F_n}  \ .\cr} $$
where $F\cdot J=\sum_a F^a J^a$ and
$F^a$ is an arbitrary set of real coefficients. In the second line
$F_n$ denotes $F\cdot J $ evaluated
in the matrix representation of the $n^{\rm th}$ level.

Theories that do not have any spin-1 operators fall outside the scope
of this paper. So far only one such theory is known, the so-called
``monster module'', which can be obtained as a modular invariant of
the $\Zbf_2$-twisted Leech lattice CFT, and which is conjectured to
be the only $c=24$ meromorphic theory without spin-1 operators.
In all other cases one may
group all spin-1 operators into the adjoint representations of some
semi-simple Lie-algebra, plus possibly some additional $U(1)$ factors.

As in the case of anomaly cancellation, we are interested in the
modular transformation properties of the character valued partition
function. They follows directly from the fact that, for $F=0$, the
partition function is modular invariant, \ie\
$$ P\left({a \tau + b \over  c \tau + d},0\right)
= P (\tau, 0),\ \ \ \ \ a,b,c,d \in \Zbf,
\ \ ad-bc =1 \ , $$
where, for convenience we have traded the variable $q$ for $\tau$,
with $q=e^{2\pi i \tau}$. If we focus on one Kac-Moody algebra or
$U(1)$ factor, we can express $P$ in terms of the characters
${\cal X}_i(\tau)$ of the
representations of that algebra
$$ P(\tau,0)  = \sum_i {\cal X}_i(\tau) P'_i(\tau) \ ,\eqn\Split $$
where $P'_i$ is a character of the ``unknown'' part of the theory.

If we choose $F$ entirely within the Kac-Moody or $U(1)$ factor under
consideration, then the dependence on $F$ enters only through the
characters ${\cal X}_i$. The modular transformation properties of
``character valued'' Kac-Moody characters is known. Under the two
generating transformations of the modular group they transform as follows
\KaPe, \Gin
$$ \eqalign {\tau \rightarrow \tau+1 \ &: \ \ \
  {\cal X}_i(\tau+1,F)=e^{2\pi i (h_i - c/24)  } {\cal X}_i(\tau,F) \cr
 \tau \rightarrow -{1\over \tau} \ &: \ \ \
  {\cal X}_i(-{1\over\tau}, {F\over\tau})=
e^{-{i\over 8 \pi  \tau}
{k\over g} \Tr F^2 }S_{ij}
{\cal X}_j(\tau,F)\ . \cr }\eqn\Trafo$$
Here $g$ is the dual Coxeter number of the Kac-Moody algebra.
The normalization of the Kac-Moody
generators is defined by \KacMoody\ plus the requirement that the
smallest allowed value for $k$
must be 1. The Lie-algebra generators $J^a_0$ are taken to be hermitean,
and the real, anti-symmetric
structure constants $f_{abc}$
satisfy $f_{abc}f_{abe}=2 g \delta_{ce}$.
In \Trafo\
the representation in which the trace is evaluated is the adjoint
representation.
Any other non-trivial representation may be used,
provided one changes the factor multiplying the trace (for example,
to compare with \ScW\ one should use the vector representation of
$SO(2N)$, and compensate the change in normalization by omitting
the factor ${1\over g}$; furthermore
one should replace $F$ by $iF/2\pi$). For $U(1)$ factors
the adjoint representation cannot be used, since it is trivial.
Instead, any representation
with a non-vanishing quadratic trace can be used, in which case one
must replace ${k\over g}\Tr F^2$ in \Trafo\ by $N\Tr F^2$.
Although
it is not difficult to obtain
the correct normalization factor $N$, for our purposes it
turns out to be irrelevant. In the following all results will be
presented explicitly for non-abelian factors only, but it is
straightforward to make the appropriate changes for abelian ones.
the appropriate changes for abelian ones.

Because $P(\tau,0)$ is modular invariant,
the phase $e^{2\pi i (h_i-c/24)}$
and the matrix $S_{ij}$ are compensated  by the transformation of
$P'$. From this we can deduce how $P(\tau,F)$ transforms
$$ P\left({a\tau+b\over c\tau+d},{F\over c\tau+d}\right) =
  \exp[{{-ic\over 8 \pi (c\tau+d)}\sum_{\ell}
  {k_{\ell}\over g_{\ell}}
\Tr F_{\ell}^2 }]\  P(\tau,F)  \ .\eqn\TrafoTwo  $$
Here we have allowed for the possibility that $F$ has components
$F_{\ell}$ in several simple factors of the Kac-Moody algebra.
To simplify the notation we define
$$ {\cal F}^2 \equiv \sum_{\ell } {k_{\ell}\over g_{\ell}}
                                  \Tr F_{\ell}^2 \ . $$

Now we expand $P(q,F)$ in traces of $F$, and make use of the
theory of modular functions to constrain the coefficients, which are
functions of $q$. A few facts about modular functions will be
mentioned below, but for more details see \ScW, or references
therein.
In the absence of the
exponential factor in \TrafoTwo, the coefficient function of a trace
of order $n$ is a modular function of weight $n$.
Such functions can
all be expressed in terms of the Eisenstein functions $E_4$ and $E_6$,
$$\eqalign{  E_4(q)&=1+240\sum_{n=1}^{\infty}
             { n^3 q^n \over 1 -  q^n } \ , \cr
             E_6(q)&=1-504\sum_{n=1}^{\infty}
             { n^5 q^n \over 1 -  q^n } \ , \cr} $$
which have weight 4 and 6 respectively. Any entire
modular function (which, as a function of $\tau$, has
no poles
in the complex upper half-plane including the point $\tau=i\infty$)
has non-negative and even
modular weight, and
can be written as a polynomial
in $E_4$ and $E_6$. The number of parameters in such a polynomial
is $k$ for weight $12k$, as well as
for weights  $12k+4$, $12k+6$, $12k+8$ and $12k+10$, and $k-1$
for weights $12k+2$.

The coefficient functions appearing in the expansion of
$c=24$ partition function
are in general not entire functions, because of the ``tachyon'' pole
at $q=0$ ($\tau=i\infty$).
To obtain
modular functions with poles at $q=0$ one defines first
$$ \eqalign{
\Delta(q)&={1\over 1728}\left((E_4)^3- (E_6)^2 \right)\cr
         &= \eta(q)^{-24}\  . \cr } $$
The most general modular function with a single pole at $q=0$
and none elsewhere in the  $\tau$ upper half-plane is a combination
of the form $(G_4)^m (G_6)^n / \Delta $, which has weight
$4m+6n-12$. The number of different functions is for weights $12k+l$
is now $k+1$ if
$l=0,4,6,8,10$, and $k$ if $l=2$.

To use these results we need to know that the coefficient
functions do not have spurious
poles elsewhere in the positive upper
half-plane.
In general, characters ${\cal X}_i$
of unitary conformal field theories
do not
have such poles if their chiral
algebra is generated by a finite number of currents $N_J$ (each of
which may have an infinite number of modes)
because
they are bounded by ${\cal X}_i  \leq q^{h_i -c/24 +N_J/24} N_0
\eta^{-N_J}$, where $N_0$ is the ground state multiplicity.
The inequality is saturated if all $N_J$ currents act independently
without null-vectors (\ie\ as $N_J$ free bosons). Null-vectors
reduce the number of states at each level and hence reduce the value
of the left-hand side with respect to the right-hand side.
Since
$\eta^{-1}$ has no poles in the upper half plane outside
$\tau=i \infty$, ${\cal X}_i$ is well-behaved as well.
Thus we will assume that
the chiral algebra is finitely generated, or in other words
that beyond a certain spin all currents can
be expressed in terms of lower spin currents.
Then not only the
partition function $P(\tau,0)$ is free of spurious poles, but also
the coefficient functions of higher traces, since in \Split\ the
Kac-Moody characters ${\cal X}_i(F,\tau)$ are explicitly known and
well-behaved, and $P'_i(\tau)$ is free of spurious poles by the same
assumption.

The term of order zero in the  expansion in $F$
is the partition function, $P(q,0)$.
This is a modular function of
weight zero with a single pole at $\tau=i\infty$.
We conclude
that
it must be equal to
$$ P(q,0) = J(q) + N \ , $$
where $J(q)$ is the absolute modular invariant
$$ J(q)={1\over q} +  196884 q + \ldots\ ,  $$
and $N$ is the number of spin-1 operators in the theory. This is
not determined by the requirement of modular invariance of the
partition function.

The exponential factor in \TrafoTwo\ is cancelled if instead of
$P(q,F)$ we consider
$$ \tilde P(q,F)= e^{-{1\over48} E_2(q) {\cal F}^2  }
            P(q,F) \ ,\eqn\CorrFac $$
where
$$           E_2(q) =1-24 \sum_{n=1}^{\infty}
             { n q^n \over 1 -  q^n }  \ . $$
This function transforms as a modular function of weight 2, plus
an ``anomalous'' extra term,
$$ E_2\left({a\tau+b\over  c\tau+d}\right)=
   (c\tau+d)^2E_2(\tau) - {6i\over\pi} c (c\tau+d) \ . $$
Now we
can apply the foregoing arguments to $\tilde P(q,F)$, and from
the expansion of $\tilde P$ derive that of $P$ by multiplying
with the inverse of the correction factor in \CorrFac. An example
of such a computation
(for the $O(32)$ heterotic superstring) can be found in
appendix C of \LNSW.

The resulting expression has many undetermined parameters, some of
which can be determined from our knowledge of the zeroth and first level
of the theory. Since the ground state is a singlet representation, all
coefficient functions of non-trivial traces must be free of poles.
This condition leads to the following parametrization of $P(q,F)$
$$ \eqalign{
P(q,F) &= \exp({1\over  48} E_2(q) {\cal F}^2
                  ) \cr
&\times \{ J(q) + N  + {(E_4)^3 \over \Delta }
  [ \cosh({1\over48} \sqrt{E_4(q)} {\cal F}^2 ) - 1]\cr
 &\ \ - \sqrt{E_4(q)} E_4(q) E_6(q)
  [ \sinh({1\over48} \sqrt{E_4(q)} {\cal F}^2 )  ] \cr
+ &\sum_{n=4}^{\infty} E_{n}a_{n,i} {\cal F}^n_i\ \}  .
 \cr  }\eqn\Traces$$
Here ${\cal F}^n_i$ denotes traces of $n^{\rm th}$ order in $F$.
In general there are many possible combinations
of traces of a given order (\eg\ at fourth order one can have
$\Tr F^4, (\Tr F^2)^2$ or $\Tr( F_1)^2 \Tr(F_2)^2$),
and the different types are
labelled  by $i$. Each different trace appears with a parameter
$a_{n,i}$. The functions $E_n$ are just the Eisenstein functions
or polynomials of such functions, normalized to have a first
coefficient equal to 1:
$E_8=(E_4)^2$, $E_{10} = E_4
E_6$, $E_{12}=\alpha (E_4)^3 + (1-\alpha) (E_6)^2$, etc. Note that
extra parameters start appearing in the definition of $E_{12}$.
The purpose of the ``cosh'' and ``sinh'' terms is simply to
cancel the ${1\over q}$ poles in coefficient functions of
combinations of quadratic traces.

Now compare this expression with the exactly known expansion of
$P(q,F)$ to the first level:
$$ P(q,F) = {1\over q} + \sum_{\ell}\Tr e^{F_{\ell}}
                  + \hbox{higher order terms in q} \ . $$
Here all traces must be evaluated in the representations of the first
level, \ie\ the adjoint representations of the Kac-Moody and $U(1)$
factors. Comparing the quadratic traces in both expressions we find
immediately
$$ {k_{\ell} \over g_{\ell}}({N\over 48} - {1\over2})= {1\over 2} \ ,
\eqn\MiracleOne$$
for non-abelian factors, and
$$ N_{\ell} ({N\over 48} - {1\over2})= 0 \eqn\MiracleTwo   $$
for $U(1)$ factors, because then $F_{\ell}$ is trivial in the adjoint
representation.

This leads first of all
to the conclusion that the value of ${g\over k}$ must be the same
for every Kac-Moody algebra appearing in a given $c=24$ theory, and
furthermore that abelian and non-abelian factors cannot be mixed
(this is already known to be true for the Niemeier lattices, see \eg\
\GoOl). Secondly,
we learn that the value of ${g\over k}$ is determined
by the total number of spin-1 operators. If $N=24$ only $U(1)$ factors
are allowed, and then we have no choice but to saturate $N$ entirely
with generators of $U(1)$'s \ie\ free bosons. The resulting theory
falls thus within the classification of Niemeier, and is in fact
the conformal field theory of the Leech lattice. But also for larger
values of $N$ we arrive at an interesting conclusion. The total
central charge of the Kac-Moody system is
$$ c_{K\!M}=\sum_{\ell} {k_{\ell}
\dim_{\ell} \over k_{\ell} + g_{\ell}}\ ,$$
where $\dim_{\ell}$ is the dimension of the adjoint representation
of the ${\ell}^{\rm th}$ factor. Substituting \MiracleOne\ we find
$$ c_{K\!M}={24\over N} \sum_{\ell} \dim_{\ell} = 24 \ , $$
so that the entire central charge is saturated by the
Kac-Moody part of the theory. Hence we conclude that
if a $c=24$ theory has any
spin-1 operators, it is either the Leech lattice CFT or can be
entirely written as a tensor product of Kac-Moody algebras.

It is now straightforward to obtain a complete list of all Kac-Moody
algebras that can appear. In total there are 221 combinations (this
does not
include $N=0$ (no spin-1 algebra) and $N=24$ ($U(1)^{24}$)). This list
contains of course all 39 presently known cases: the 23 algebras of the
Niemeier lattices \Nie, the 14 additional ones
obtained from the $\Zbf_2$-twisted Niemeier lattices
\doubref\BB\DGM\rlap,\foot{One combination listed in \DGM, namely
$(A_3)^4 (A_2)^2$, is incorrect, and should be
$(A_{3})^4 (A_1)^4$.}and the 2 examples
of \ScYc.

The next
task is to try and find modular invariant combinations of the
characters of each combination of Kac-Moody algebras.
Of the  221 combinations 24 require no further discussion because
the only valid character combinations are those of the Niemeier
lattices\rlap.\foot{In principle only even self-dual lattices have
been classified in \Nie, and this is not necessarily the same
as classifying all modular invariants.
There might exist character
combinations that do not have a lattice interpretation, for example
because there are multiplicities higher than 1. However, all primary
fields
of level-1 simply-laced Kac-Moody algebras are simple currents. All
closed chiral algebra that can be built out of simple currents and that
can appear in modular invariant partitions have been classified in
\BeBT. This rules out any ``non-lattice'' solutions.}For
the other 17 Kac-Moody combinations for which modular invariants
are known this is not quite true, because there might still exist other
modular invariant combinations.

Modular invariant combinations of characters will certainly not
exist for all 221 combinations.
This can especially not
expected to be true for small $N$, where
often many algebras with small dimensions occur, and accidental
solutions to the condition $\sum_{\ell} \dim_{\ell}=N$
are likely. One way to eliminate such accidental combinations is to
use the other trace identities one can obtain from \Traces. If
the functions $E_n$ have no free parameters themselves, then
the coefficients $a_{n,i}$ can be determined at the first level. One
then gets predictions for traces of representations at the second
level.
If there are no representations with conformal spin 2 that
can satisfy these identities, then there cannot exist a meromorphic
modular invariant. If there is a solution with positive integer
coefficients one can either try to go on to the third level, or
directly use the matrix $S$ to determine the remaining unknown
coefficients, and prove that indeed one gets an invariant. This will
in any case have to be done to be certain that one has indeed a
solution.

In general, the total number of spin-2 representations is very large,
but so is the number of trace identities. The traces ${\cal F}^n_i$
of the second level representations are completely determined by
modular invariance
for $n=0,2,4,6,8,10$ and $n=14$ (odd traces must
vanish).
If there are
several Kac-Moody factors there are of course mixed traces with
components in more than one factor. So far our analysis is based
only upon traces that are of at most order 2 in each factor. This
eliminates at least 90 of the initial 221 combinations. For
another 24 combinations a solution to the trace identities exists (not
including the 39 already known solutions).
Although {\it a priori} the coefficients of each representation are
allowed  to be any positive integer, in these solutions they are
(up to unresolved ambiguities) either 0 or 1, which is a strong
indication that these solutions are not accidental. Of course, further
checks are required. Finally, for the remaining
combinations the existence
of solutions is still
unclear, because the number of fields is too large
in comparison with the number of trace identities. The use of higher
order traces might improve this very significantly. In any case, the
classification of $c=24$ theories (or at least the corresponding
modular invariant combinations of Kac-Moody algebras)
with at least one spin-1 current is now
reduced to a finite algorithm.

The values of $N$ for which $c=24$ theories may exist are 0, 24-28,30,
32,36,40,42,48,56, all multiples of 12 from 60 to 168, all multiples of
24 from 192 to 408, 300, 456,552,624,744 and 1128. For all
combinations with $N\geq 300$ solutions are already known. A new
modular invariant
appears for $N=288$ with a Kac-Moody algebra
$B_{6,1} C_{10,1}$. This invariant can easily be understood in the
following way. Because of rank-level duality for the $C_{n,k}$ algebras,
the well-known invariant of $SU(2)$ level 10 (\ie\ $C_{1,10}$) implies
the existence of a similar invariant for $C_{10,1}$ \Ver.
The latter has
three primary fields which turn out to have the fusion rules of
the Ising model, or of $B_{6,1}$. Glueing these theories together in
the obvious way yields a meromorphic $c=24$ invariant.

A second
new example is $A_{5,1} \times E_{7,3}$, obtained by glueing the
characters of the first factor to those of a non-diagonal
modular invariant of the second. This $E_{7,3}$ invariant is
apparently not yet known, and has the form
$$ |{\cal X}_{0000000} +{\cal X}_{0000011}|^2
  + | {\cal X}_{0000001} + {\cal X}_{0000030} |^2
  + 2 | {\cal X}_{0000100} |^2 + 2| {\cal X}_{1000010} |^2 \ . $$
After resolving the two fixed points one obtains a total of six
fields, with the same fusion rules as $A_{5,1}$. The rest of the
discussion is completely analogous as that of the
$A_{2,2} \times F_{4,6}$ example of \ScYc. Note that in all these
cases one discovers extensions of the chiral algebra (of $C_{10,1}$,
$E_{7,3}$ and $F_{4,6}$ respectively) by operators with  spin larger
that one, since we have already accounted for all spin-1 operators
by satisfying the sum rule
$\sum_{\ell} \dim_{\ell} = N $ at the first level. Therefore one
always finds extension
that can {\it not}
be obtained from conformal embeddings.
(This should not be
confused with  the fact that Kac-Moody algebras of
all examples of \DGM\ can
be conformally embedded in the Kac-Moody algebras of
Niemeier theories. In those cases the
extra spin-1 currents have been removed by the $\Zbf_2$ twist).
These two examples are
only new solutions of the 19 candidates mentioned above for which
modular invariance has been checked so far, but there are several
others that should not be too difficult to verify.

A modular invariant combination of Kac-Moody characters is not yet
a conformal field theory, but it is an important step in that
direction. The next step could be either to find an explicit
realization, or to try to compute correlation functions or
higher genus partition functions. It should be emphasized that
there is a crucial difference between
knowing the field content of a $c=24$ CFT in terms of representations
of  a $c=24$ Kac-Moody sub-algebra of the full chiral algebra
and just
knowing the partition function $P(q,0)$. This point was apparently
not appreciated by the authors of \DGM, who remark that one
could not distinguish the $E_8 \times E_8$ and $O(32)$ theories in this
way. One certainly cannot distinguish those theories if
one only knows $P(q,0)$, but if one knows $P(q,F)$ and
the two different ways of
writing it in terms of characters of $E_8\times E_8$ or $O(32)$
respectively,
then one has certainly made the distinction.
The point is that $P(q,0)$ contains only information about
Virasoro representations, and the Virasoro algebra is too small to
determine correlation functions except when $c<1$. If $P(q,0)$ is
all one knows about a $c=24$ theory one cannot even begin to compute
operator products or correlators.
On the other
hand, if the full $c=24$ theory is covered by a combination of
Kac-Moody algebras, one already knows a large enough sub-algebra of the
chiral algebra to determine correlators and operator products, although
the actual computations become extremely tedious due to
the fact that one has an off-diagonal invariant
with a (known!) extension of the
chiral algebra. Inconsistencies and ambiguities in these
computations are not {\it a priori} ruled out.
In any case, since the
number of allowed Kac-Moody combinations is only 221, a good
strategy is to try first to find modular invariant character
combinations, and worry about further consistency checks later. This
is certainly sufficient for the classification of $d=10$
heterotic strings, to which we return now.

All possible Kac-Moody algebras in which $D_{8,1}$ can be embedded occur
for $N\geq360$. Since no new theories appeared for $N\geq 300$ it follows
that there are no more ten-dimensional heterotic strings than
the already known ones (for the only non-lattice theory,
$B_{8,1} E_{8,2}$, one has to verify that there is no other modular
invariant, but this is trivial). Thus the main goal of this paper
has been accomplished.

In a forthcoming paper the classification of the remaining $c=24$
theories will be discussed, and, depending on the complexity of the
calculations and the available computer time, a more or less
complete classification of the modular invariants will be presented.

\ack
I would like to thank Eliezer Rabinovici
for discussions,
Jurgen Fuchs for providing
the
$S$-matrices of $C_{10,1}$ and $E_{7,3}$, and
the theory group of NIKHEF (Amsterdam), where most of this
work was done, for hospitality.

\par \penalty-4000\vskip\chapterskip
   \spacecheck\referenceminspace \immediate\closeout\referencewrite
   \referenceopenfalse
   \line{\fourteenrm\hfil REFERENCES\hfil}\vskip\headskip
   \endlinechar=-1
   \input referenc.texauxil
   \endlinechar=13
   \bye